\documentclass[journal,9pt]{IEEEtran}

\usepackage{amsmath,amsfonts}
\usepackage{algorithmic}
\usepackage{algorithm}
\usepackage{array}

\usepackage[caption=false,font=normalsize,labelfont=sf,textfont=sf]{subfig}

\usepackage{textcomp}
\usepackage{stfloats}
\usepackage{url}
\usepackage{verbatim}
\usepackage{graphicx}
\usepackage{cite}
\usepackage{hyperref}
\begin{document}

\title{\huge{General Method for Conversion Between Multimode Network Parameters}}

\author{\Large{Alexander Zhuravlev and Juan D. Baena,~\IEEEmembership{Member,~IEEE}}

\thanks{Copyright was transferred to IEEE.}

\thanks{This paper was produced by the IEEE Publication Technology Group. They are in Piscataway, NJ.}

\thanks{This work was supported by the Russian Science Foundation (Project No. 24-45-02020).}

\thanks{A. Zhuravlev is with the School of Physics and Engineering, ITMO University, St. Petersburg, Russia (e-mail: a.zhuravlev@metalab.ifmo.ru).}
\thanks{J. D. Baena is with the Department of Physics, Universidad Nacional de Colombia, Bogota 111321, Colombia (e-mail: jdbaenad@unal.edu.co).}}

\markboth{Preprint}%
{Shell \MakeLowercase{\textit{et al.}}: A Sample Article Using IEEEtran.cls for IEEE Journals}

\IEEEpubid{0000--0000~\copyright~2023 IEEE}

\maketitle

\begin{abstract}
Different types of network parameters have been used in electronics since long ago. The most typical network parameters, but not the only ones, are $S$, $T$, $ABCD$, $Z$, $Y$, and $h$ that relate input and output signals in different ways. There exist practical formulas for conversion between them. Due to the development of powerful software tools that can deal efficiently and accurately with higher-order modes in each port, researchers need conversion rules between multimode network parameters. However, the usual way to get each conversion rule is just developing cumbersome algebraic manipulations which, at the end, are useful only for some specific conversion. Here, we propose a general algebraic method to obtain any conversion rule between different multimode network parameters. It is based on the assumption of a state vector space and each conversion rule between network parameters can be interpreted as a simple change of basis. This procedure explains any conversion between multimode network parameters under the same algebraic steps.
\end{abstract}

\vskip0.5\baselineskip

\section{Introduction}

Microwave devices play a crucial role in modern telecommunication systems, because they provide necessary manipulations with fields to receive and process a signal: filtering, beamforming, polarization control, amplifying, and signal mixing, among others. To model and measure the performance of a microwave device, engineers and researchers usually use network parameters that connect fields (electric and magnetic fields) or certain integrals of them (effective voltage and current) at the terminals of the device. Among the most used network parameters used in practice, there are the scattering $S$, transfer $T$, chain $ABCD$, impedance $Z$, and admittance $Y$ parameters \cite{Collin1992, Pozar2012}. For microwave networks, the scattering parameters $S$ are commonly used and measurable using any commercial vector network analyzer. In antenna engineering, the impedance $Z$ and admittance $Y$ network parameters might be preferred to control the input impedance of an antenna. In addition, they can be calculated using the induced electromotive force method presented in \cite{Otto1969}. For an arrangement of several connected networks, the $ABCD$ matrix or the $T$ matrix can be very useful because the parameters of all parts can be multiplied in the cascade to obtain the matrix of the overall structure. Then the resulting $T$ or $ABCD$ matrices can be converted to any other type of network parameters. Although the parameters $T$ and $ABCD$ are usually difficult to measure, they can be derived from measured $S$ parameters. By the way, $T$ and $ABCD$ network parameters are usually used to calibrate and remove the effects of cables and connectors that connect the device under test to the vector network analyzer.   

\begin{figure*}
\centering
\includegraphics[width=\textwidth]{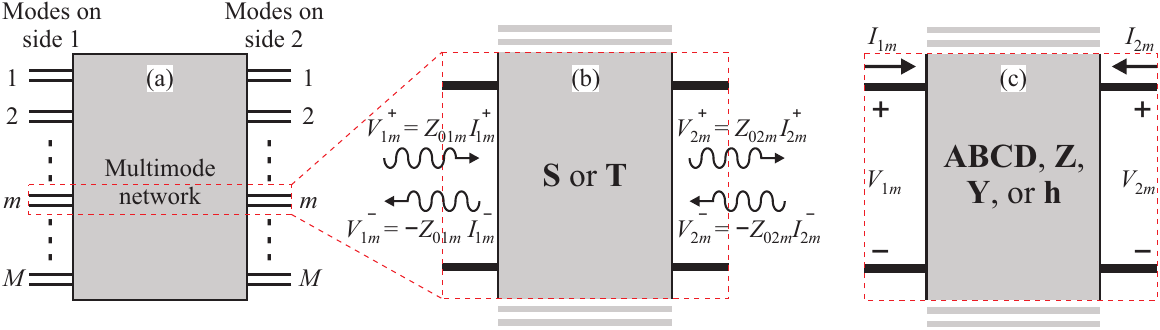}
\caption{Diagram of a multimode network. In (a) a general view shows $M$ modes attached on each side, while (b) and (c) only show the portion containing the $m$-th modes for a better visualization of voltages and currents used in definitions of network parameters.}
\label{fig:network}
\end{figure*}

We can conclude that conversion rules between different network parameters play an important role in engineering and research practice. Usually, tables of practical conversion rules are presented in microwave engineering textbooks \cite{Pozar2012} or research papers \cite{Frickey1994}. However, they consider monomode 2-port network parameters and thus ignore the effect of higher-order modes. For example, it may be important for densely packed structures such as those studied in \cite{Amari2000, Bandlow2008, Bongard2009, Martini2014, Wong2016, Mesa2018, Bagheriasl2019, Mesa2021, Giusti2022, Escobar2023, Giusti2024}. It is worth noting that a multimode network can also be treated as a multiport network and vice versa because each mode can be associated with some transmission line with its own characteristic impedance. Some conversion rules were presented in \cite{Frei2008} for $S \leftrightarrow T$ and \cite{Ma2022} for $S \leftrightarrow ABCD$. Finally, the work \cite{Reveyrand2018} extends all the conversion rules of \cite{Frickey1994} to the multiport case. However, this paper contains the final formulas but no general algorithm that could facilitate conversion between others. Our work suggests a simple algorithm to convert multimode network parameters that can be easily extended to any possible conversion between multimode network parameters.

\section{Definitions of multimode network parameters}

Consider a network connected to multiple modes from each side as sketched in Fig.~\ref{fig:network}(a). The voltage amplitudes of the incoming / outgoing signals are shown in Fig.~\ref{fig:network}(b). It is convenient to arrange them together in the following vectors: $ \textbf{V}_s^\pm =\left[V_{s1}^\pm \hdots V_{sm}^\pm   \hdots V_{sM}^\pm\right]^{\text{T}}$, where $s = 1,2$ indicates one side of the network, the superscript $+(-)$ refers to incoming (outgoing) signals, $m$ is the mode, and $M$ the total number of modes. Alternatively, the system can also be described using the net voltages and currents at the edge of the network for each mode, as shown in Fig.~\ref{fig:network}(c), where 
\begin{equation}
\begin{gathered}
    \textbf{V}_s = \textbf{V}_s^+ + \textbf{V}_s^- \\
    \textbf{I}_s = \textbf{Y}_s \cdot (\textbf{V}_s^+ - \textbf{V}_s^-)
\end{gathered}
\label{eq:VI}
\end{equation}
Here $\textbf{Y}_s = \text{Diag}[Y_{0s1}, \dots Y_{0sm}, \dots Y_{0sM}]$ is a diagonal matrix containing the mode admittances of all modes from the side $s$. In other words, $V_{sm} = V_{sm}^+ + V_{sm}^-$ and $I_{sm} = Y_{0sm}(V_{sm}^+ - V_{sm}^-)$ for each mode. By combining the two equations in \eqref{eq:VI}, it can also be written as 
\begin{equation}
    \textbf{V}_s^\pm = \frac{1}{2} (\textbf{V}_s \pm \textbf{Z}_s \cdot \textbf{I}_s )
\label{eq:V}
\end{equation}
where $\textbf{Z}_s = \textbf{Y}_s^{-1}$ is just a diagonal matrix formed with the mode impedances. Now, using the voltages and currents defined in Fig.~\ref{fig:network}, several multimode network parameters can be defined as follows: 
\begin{itemize}
\item scattering matrix \textbf{S}
\begin{equation}
    \begin{bmatrix}
         \textbf{V}_1^- \\
         \textbf{V}_2^-
    \end{bmatrix} =
    \underbrace{\begin{bmatrix}
        \textbf{S}_{11} & \textbf{S}_{12} \\
        \textbf{S}_{21} & \textbf{S}_{22}
    \end{bmatrix}}_\textbf{\large{S}}
    \cdot
    \begin{bmatrix}
         \textbf{V}_1^+ \\
         \textbf{V}_2^+
    \end{bmatrix}
\label{eq:S}
\end{equation}
\item transfer matrix \textbf{T}
\begin{equation}
    \begin{bmatrix}
         \textbf{V}_1^+ \\
         \textbf{V}_1^-
    \end{bmatrix} =
    \underbrace{\begin{bmatrix}
         \textbf{T}_{11} & \textbf{T}_{12} \\
         \textbf{T}_{21} & \textbf{T}_{22}
    \end{bmatrix}}_\textbf{\large{T}}
    \cdot
    \begin{bmatrix}
         \textbf{V}_2^- \\
         \textbf{V}_2^+
    \end{bmatrix}
\label{eq:T}
\end{equation}
\item chain matrix \textbf{ABCD}
\begin{equation}
    \begin{bmatrix}
         \textbf{V}_1 \\
         \textbf{I}_1
    \end{bmatrix} =
    \underbrace{\begin{bmatrix}
         \textbf{A} & \textbf{B} \\
         \textbf{C} & \textbf{D}
    \end{bmatrix}}_\textbf{\large{ABCD}}
    \cdot
    \begin{bmatrix}
         \textbf{V}_2 \\
         -\textbf{I}_2
    \end{bmatrix}
\label{eq:ABCD}
\end{equation}
\item impedance matrix \textbf{Z}
\begin{equation}
    \begin{bmatrix}
         \textbf{V}_1 \\
         \textbf{V}_2
    \end{bmatrix} =
    \underbrace{\begin{bmatrix}
         \textbf{Z}_{11} & \textbf{Z}_{12} \\
         \textbf{Z}_{21} & \textbf{Z}_{22}
    \end{bmatrix}}_\textbf{\large{Z}}
    \cdot
    \begin{bmatrix}
         \textbf{I}_1 \\
         \textbf{I}_2
    \end{bmatrix}
\label{eq:Z}
\end{equation}
\item admittance matrix \textbf{Y}
\begin{equation}
    \begin{bmatrix}
         \textbf{I}_1 \\
         \textbf{I}_2
    \end{bmatrix} =
    \underbrace{\begin{bmatrix}
         \textbf{Y}_{11} & \textbf{Y}_{12} \\
         \textbf{Y}_{21} & \textbf{Y}_{22}
    \end{bmatrix}}_\textbf{\large{Y}}
    \cdot
    \begin{bmatrix}
         \textbf{V}_1 \\
         \textbf{V}_2
    \end{bmatrix}
\label{eq:Y}
\end{equation}
\item hybrid matrix \textbf{h}
\begin{equation}
    \begin{bmatrix}
         \textbf{V}_1 \\
         \textbf{I}_2
    \end{bmatrix} =
    \underbrace{\begin{bmatrix}
         \textbf{h}_{11} & \textbf{h}_{12} \\
         \textbf{h}_{21} & \textbf{h}_{22}
    \end{bmatrix}}_\textbf{\large{h}}
    \cdot
    \begin{bmatrix}
         \textbf{I}_1 \\
         \textbf{V}_2
    \end{bmatrix}
\label{eq:h}
\end{equation}
\end{itemize}
It should be noted that these definitions are similar to the standard 2-port network parameters \cite{Frickey1994} and are direct extensions for the study of multimode networks. Basically, each scalar parameter has been replaced with a matrix parameter ($\textbf{S}_{ij}$, $\textbf{T}_{ij}$, $\textbf{A}, \hdots$) of size $M \times M$, being $M$ the number of modes. In fact, to highlight their matrix nature, they are written in bold here. Therefore, the multimode network parameters ($\textbf{S}$, $\textbf{T}$, $\textbf{ABCD}, \hdots$) have size $2M \times 2M$. 

\begin{table}
    \centering
        \caption{Coordinate systems of the state space suitable for each type of network parameters}
    \begin{tabular}{c}
        \includegraphics[width=\columnwidth]{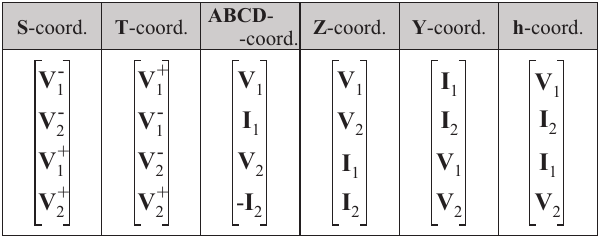} \\
    \end{tabular}
    \label{tab:coordinates}
\end{table}

\begin{table*}
    \centering
        \caption{Bases of the state vector space expressed in the coordinate systems defined in Table~\ref{tab:coordinates}}
    \begin{tabular}{c}
        \includegraphics[width=\textwidth]{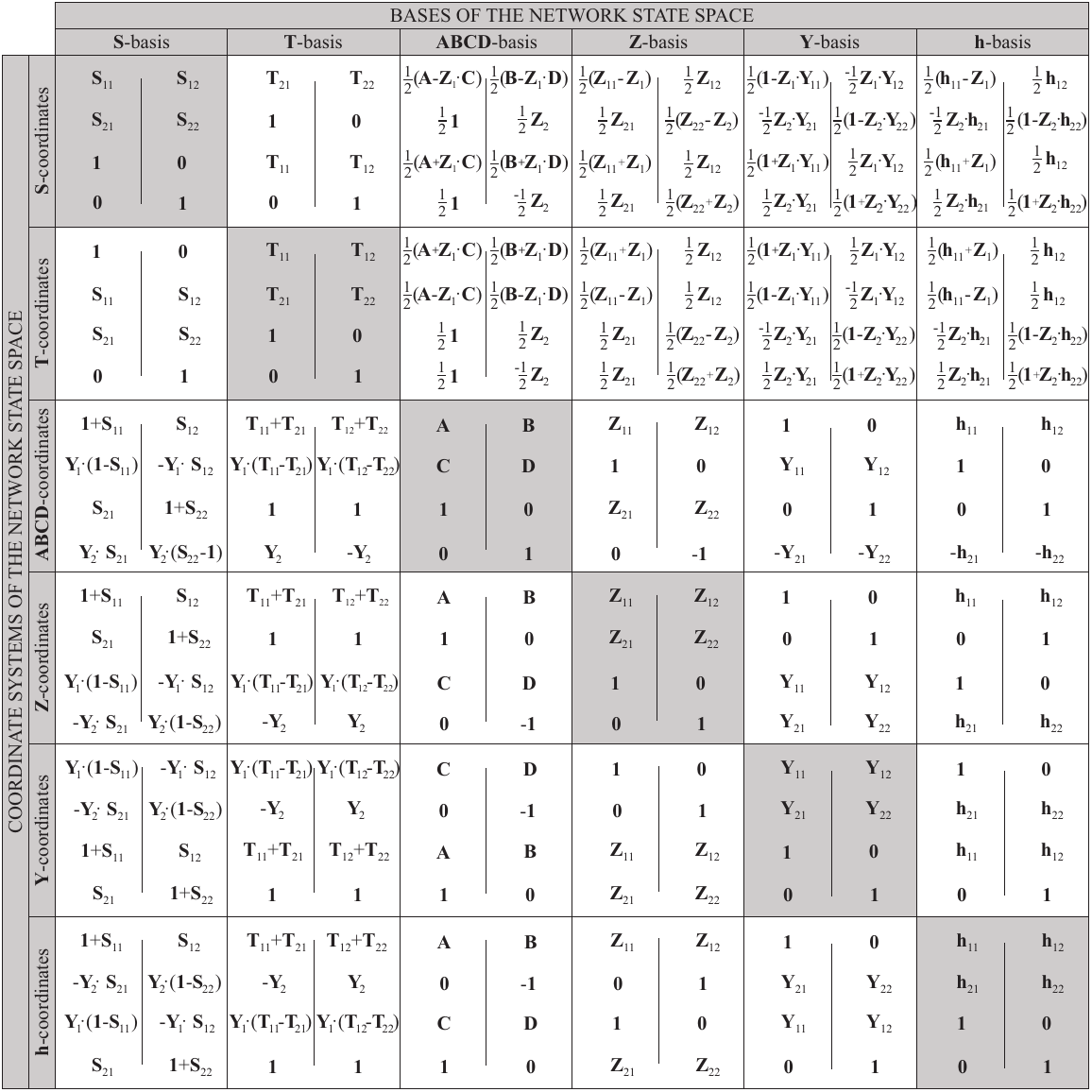} \\
    \end{tabular}
    \label{tab:bases}
\end{table*}

\section{State vector space of the network}

The state of a network corresponds to the full information of the incoming and outgoing signals. Once a network has been specified, the outgoing signals depend only on the incoming signals, and thus the dimension of the state vector space is just $2M$. In other words, only half of the state coordinates are really free. In fact, it could be said that it is the reason why all previous definitions of network parameters have size $2M \times 2M$. Taking into account the standard network matrices introduced in (\ref{eq:S}--\ref{eq:h}), it is natural to define the coordinate systems contained in Table~\ref{tab:coordinates}. The vectors shown in this table were all made by staking in a single column the two vectors around each network matrix, so that each contains $4M$ complex components. 

As usual, a basis of the state space vector could be defined. Table~\ref{tab:bases} shows six possible bases of the state space, each suitable for each type of network matrix. All places in the same column of the table contain the same basis, but expressed in different coordinate systems. Let us explain how this table is made. In a first stage, the bases of the diagonal places of the table (shadowed places) were arbitrarily filled using zeros and ones in the two lower components that correspond to the vector on the right side of equations (\ref{eq:S}--\ref{eq:h}). The upper half can then be filled using the same equations in such a way that it shows the corresponding network parameter. It should be noted that the two columns of each cell of the table actually represent $2M$ columns (one for each basis vector), as well as the bold style $\textbf{0}$ and $\textbf{1}$ are the zero matrix and the identity matrix of size $M \times M$. In a second stage, using the definitions of coordinate systems in Table~\ref{tab:bases} and the transformations \eqref{eq:VI} and \eqref{eq:V}, the six previous bases were rewritten in all coordinate systems. It is not worth adding that this table could easily be extended if some new network matrix was defined.

\section{Conversion between two network matrices}

The last stage, and goal of this paper, is to get the conversion rules. Consider any couple of two multimode network parameters $\textbf{M}$ and $\textbf{N}$, which have been chosen among $\textbf{S}$, $\textbf{T}$, $\textbf{ABCD}$, $\textbf{Z}$, $\textbf{Y}$, and $\textbf{h}$, or whatever new definition. Since each basis vector in Table~\ref{tab:bases} must satisfy all network matrices, it is possible to write
\begin{equation}
    \begin{bmatrix}
         \text{Upper half of} \\
         \text{$\textbf{M}$-basis expressed} \\
         \text{in $\textbf{N}$-coordinates}
    \end{bmatrix}     
    = \textbf{N} \cdot
    \begin{bmatrix}
         \text{Lower half of} \\
         \text{$\textbf{M}$-basis expressed} \\
         \text{in $\textbf{N}$-coordinates}
    \end{bmatrix}
\end{equation}
where the upper and lower halves mean the first two rows and the last two rows of the matrices given in Table~\ref{tab:bases}, respectively. From this, the general formula for converting from $\textbf{M}$ to $\textbf{N}$ is the following:
\begin{equation}
    \textbf{N} =
    \begin{bmatrix}
         \text{Upper half of} \\
         \text{$\textbf{M}$-basis expressed} \\
         \text{in $\textbf{N}$-coordinates}
    \end{bmatrix}     
    \cdot
    \begin{bmatrix}
         \text{Lower half of} \\
         \text{$\textbf{M}$-basis expressed} \\
         \text{in $\textbf{N}$-coordinates}
    \end{bmatrix}^{-1}
\label{eq:MtoN}
\end{equation}
For example, using the \textbf{S}-basis expressed in \textbf{T}-CS shown in Table~\ref{tab:bases} (2nd row, 1st column), we can write the conversion rule from $\textbf{S}$ to $\textbf{T}$ as follows:
\begin{equation}
    \begin{bmatrix}
         \textbf{T}_{11} & \textbf{T}_{12} \\
         \textbf{T}_{21} & \textbf{T}_{22}
    \end{bmatrix} =
    \begin{bmatrix}
         \textbf{1}      & \textbf{0} \\
         \textbf{S}_{11} & \textbf{S}_{12}
    \end{bmatrix}    
    \cdot
    \begin{bmatrix}
         \textbf{S}_{21} & \textbf{S}_{22} \\
         \textbf{0}  & \textbf{1}
    \end{bmatrix}^{-1}
\label{eq:StoT}
\end{equation}
Reciprocally, we can use the \textbf{T}-basis in \textbf{S}-CS also shown in Table~\ref{tab:coordinates} (1st row, 2nd column) to convert \textbf{T} into \textbf{S} through the following formula:
\begin{equation}
    \begin{bmatrix}
         \textbf{S}_{11} & \textbf{S}_{12} \\
         \textbf{S}_{21} & \textbf{S}_{22}
    \end{bmatrix} =
    \begin{bmatrix}
         \textbf{T}_{21} & \textbf{T}_{22} \\
         \textbf{1}      & \textbf{0}
    \end{bmatrix}    
    \cdot
    \begin{bmatrix}
         \textbf{T}_{11} & \textbf{T}_{12} \\
         \textbf{0}      & \textbf{1}
    \end{bmatrix}^{-1}
\label{eq:TtoS}
\end{equation}
Finally, any other conversion formula can be obtained in a similar fashion using \eqref{eq:MtoN} together with the corresponding basis contained in Table~\ref{tab:bases}.  

\section{Conclusion}

A general method to convert any multimode network matrix into another has been presented. The method is based on the assumption of a network state vector space and the setting of several bases expressed in different coordinate systems, which have been defined in accordance with the standard definitions of network matrices. The main advantage compared to previous works is that all conversion rules can be demonstrated by following the same steps. It significantly reduces the number of algebraic manipulations and, in addition, provides very compact conversion formulas. With the aid of \eqref{eq:MtoN} together with Table~\ref{tab:bases}, it is possible to write down up to 30 different conversion rules, while previous works \cite{Frei2008, Ma2022, Reveyrand2018} presented a lower number of conversion formulas which could make it necessary to use intermediate conversions to go from one network matrix to another.

Besides, different normalizations of equivalent voltages and currents can lead to different definitions of the network matrices (S-matrix for power waves \cite{Kurokawa1965, Collin1992, Pozar2012} and  pseudo-waves \cite{Marks1992}, normalized Z-matrix \cite{Collin1992}, among
others), and thus different rules of conversion. In fact, the present work is important not only because of the practical conversion formulas shown here but also because complete demonstrations have been provided here in such a way that extensions to any other definition of network matrix could easily be incorporated following the same methodology.



\bibliographystyle{ieeetr}
\bibliography{references.bib}

\begin{thebibliography}{10}

\bibitem{Collin1992}
R.~Collin, {\em Foundations for Microwave Engineering}.
\newblock Electrical engineering series, McGraw-Hill, 1992.

\bibitem{Pozar2012}
D.~Pozar, {\em Microwave Engineering}.
\newblock Wiley, 2012.

\bibitem{Otto1969}
D.~Otto, ``A note on the induced emf method for antenna impedance,'' {\em IEEE Transactions on Antennas and Propagation}, vol.~17, no.~1, pp.~101--102, 1969.

\bibitem{Frickey1994}
D.~Frickey, ``Conversions between {S}, {Z}, {Y}, h, {ABCD}, and {T} parameters which are valid for complex source and load impedances,'' {\em IEEE Transactions on Microwave Theory and Techniques}, vol.~42, no.~2, pp.~205--211, 1994.

\bibitem{Amari2000}
S.~Amari, R.~Vahldieck, J.~Bornemann, and P.~Leuchtmann, ``Spectrum of corrugated and periodically loaded waveguides from classical matrix eigenvalues,'' {\em IEEE Trans. Microw. Theory Techn.}, vol.~48, no.~3, pp.~453--460, 2000.

\bibitem{Bandlow2008}
B.~{Bandlow}, R.~{Schuhmann}, G.~{Lubkowski}, and T.~{Weiland}, ``Analysis of single-cell modeling of periodic metamaterial structures,'' {\em IEEE Trans. Magn.}, vol.~44, no.~6, pp.~1662--1665, 2008.

\bibitem{Bongard2009}
F.~Bongard, J.~Perruisseau-Carrier, and J.~R. Mosig, ``{Enhanced Periodic Structure Analysis Based on a Multiconductor Transmission Line Model and Application to Metamaterials},'' {\em IEEE Trans. Microw. Theory Techn.}, vol.~57, no.~11, pp.~2715--2726, 2009.

\bibitem{Martini2014}
E.~{Martini}, G.~M. {Sardi}, and S.~{Maci}, ``Homogenization processes and retrieval of equivalent constitutive parameters for multisurface-metamaterials,'' {\em IEEE Trans. Antennas Propag.}, vol.~62, no.~4, pp.~2081--2092, 2014.

\bibitem{Wong2016}
J.~P.~S. Wong, A.~Epstein, and G.~V. Eleftheriades, ``Reflectionless wide-angle refracting metasurfaces,'' {\em IEEE Antennas and Wireless Propagation Letters}, vol.~15, pp.~1293--1296, 2016.

\bibitem{Mesa2018}
F.~Mesa, R.~Rodr{\'{i}}guez-Berral, and F.~Medina, ``On the computation of the dispersion diagram of symmetric one-dimensionally periodic structures,'' {\em Symmetry}, vol.~10, no.~8, p.~307, 2018.

\bibitem{Bagheriasl2019}
M.~Bagheriasl, O.~Quevedo-Teruel, and G.~Valerio, ``{Bloch analysis of artificial lines and surfaces exhibiting glide symmetry},'' {\em IEEE Trans. Microw. Theory Techn.}, vol.~67, no.~7, pp.~2618--2628, 2019.

\bibitem{Mesa2021}
F.~Mesa, G.~Valerio, R.~Rodríguez-Berral, and O.~Quevedo-Teruel, ``Simulation-assisted efficient computation of the dispersion diagram of periodic structures: A comprehensive overview with applications to filters, leaky-wave antennas and metasurfaces,'' {\em IEEE Antennas Propag. Mag.}, vol.~63, no.~5, pp.~33--45, 2021.

\bibitem{Giusti2022}
F.~Giusti, Q.~Chen, F.~Mesa, M.~Albani, and O.~Quevedo-Teruel, ``Efficient {B}loch analysis of general periodic structures with a linearized multimodal transfer-matrix approach,'' {\em IEEE Trans. Antennas Propoag.}, vol.~70, no.~7, pp.~5555--5562, 2022.

\bibitem{Escobar2023}
A.~C. Escobar, F.~Mesa, O.~Quevedo-Teruel, and J.~D. Baena, ``Homogenization of periodic structures using the multimodal transfer matrix method,'' {\em IEEE Transactions on Antennas and Propagation}, vol.~71, no.~6, pp.~4976--4989, 2023.

\bibitem{Giusti2024}
F.~Giusti, E.~Martini, S.~Maci, and M.~Albani, ``Comparison between different approaches for the design of anomalous refractors,'' {\em IEEE Transactions on Antennas and Propagation}, vol.~72, no.~4, pp.~3495--3506, 2024.

\bibitem{Frei2008}
J.~Frei, X.-D. Cai, and S.~Muller, ``Multiport {$S$}-parameter and {$T$}-parameter conversion with symmetry extension,'' {\em IEEE Transactions on Microwave Theory and Techniques}, vol.~56, no.~11, pp.~2493--2504, 2008.

\bibitem{Ma2022}
M.~Ma, F.~You, M.~Wei, T.~Qian, Y.~Chen, M.~Pan, Y.~Wang, C.~Shen, R.~Qin, T.~Wu, and S.~He, ``A generalized multiport conversion between {S} parameter and {ABCD} parameter,'' in {\em 2022 International Conference on Microwave and Millimeter Wave Technology (ICMMT)}, pp.~1--3, 2022.

\bibitem{Reveyrand2018}
T.~Reveyrand, ``Multiport conversions between {S}, {Z}, {Y}, h, {ABCD}, and {T} parameters,'' in {\em 2018 International Workshop on Integrated Nonlinear Microwave and Millimetre-wave Circuits (INMMIC)}, pp.~1--3, 2018.

\bibitem{Kurokawa1965}
K.~Kurokawa, ``Power waves and the scattering matrix,'' {\em IEEE Transactions on Microwave Theory and Techniques}, vol.~13, no.~2, pp.~194--202, 1965.

\bibitem{Marks1992}
R.~B. Marks and D.~F. Williams, ``A general waveguide circuit theory,'' {\em J. Res. Natl. Inst. Stand. Technol.}, vol.~97, pp.~533--561, 1992.

\end{thebibliography}
\nocite{*}

\end{document}